\documentclass[letterpaper,twocolumn,american,prl,aps,superscriptaddress,eqsecnum,nofootinbib]{revtex4}
\usepackage{amsfonts}%
\usepackage{amsmath}%
\usepackage{amssymb}%
\usepackage{graphicx}
\usepackage{color}

\providecommand{\U}[1]{\protect\rule{.1in}{.1in}}
\newtheorem{theorem}{Theorem}

\newtheorem{lemma}[theorem]{Lemma}

\begin{document}

\title{An information-theoretic account of the Wigner-Araki-Yanase theorem}

\author{Iman Marvian}
\affiliation{Perimeter Institute for Theoretical Physics, 31 Caroline St. N, Waterloo, \\
Ontario, Canada N2L 2Y5}
\affiliation{Institute for Quantum Computing, University of Waterloo, 200 University Ave. W, Waterloo, Ontario, Canada N2L 3G1}
\affiliation{Department of Physics and Astronomy, Center for Quantum Information Science and Technology, University of Southern California, Los Angeles, CA 90089}

\author{Robert W. Spekkens}
\affiliation{Perimeter Institute for Theoretical Physics, 31 Caroline St. N, Waterloo, \\
Ontario, Canada N2L 2Y5}

\begin{abstract}
The Wigner-Araki-Yanase (WAY) theorem can be understood as a result in the resource theory of asymmetry asserting the impossibility of perfectly simulating, via symmetric processing, the measurement of an asymmetric observable unless one has access to a state that is \emph{perfectly} asymmetric, that is, one whose orbit under the group action is a set of orthogonal states.  The simulation problem can be characterized information-theoretically by considering how well both the target observable and the resource state can provide an encoding of an element of the symmetry group.  Leveraging this information-theoretic perspective, we show that the WAY theorem is a consequence of the no-programming theorem for projective measurements. The connection allows us to clarify the conceptual content of the theorem and to deduce some interesting generalizations.
\end{abstract}
\date{Dec. 13, 2012}
\maketitle

An important question in the foundations of quantum theory is whether it is possible to devise a measurement procedure for any given Hermitian operator.  It is known that the answer can sometimes be negative, for instance, by virtue of the measurement procedure requiring superluminal communication.  Even within a nonrelativistic context, however, there can be restrictions due to conservation laws.  Specifically, the observables for which one can implement a repeatable measurement are restricted to those that commute with the conserved quantity --- this is the content of the Wigner-Araki-Yanase (WAY) theorem~\cite{Wigner1952,araki1960measurement}.
More precisely, suppose a measurement on a system $s$ proceeds by coupling $s$ to an apparatus $a$ of finite Hilbert space dimensionality via a unitary $U$ and then recording the value of a pointer observable $O_{a}$. Suppose also that $L=L_{s}\otimes I_{a} +I_{s}\otimes L_{a}$ is an additive conserved quantity, so that $\left[
L,U\right]  =0,$ and that the pointer observable satisfies $\left[  O_{a},L_{a}\right]=0$.
It then follows that the only observables $O_{s}$ on the system for which it is
possible to implement a repeatable measurement are those that satisfy $\left[
O_{s},L_{s}\right]  =0.$ \ Recent generalizations of the theorem have shown
that the condition of repeatability is not necessary; regardless of the state update rule, the only observables that
are measurable are those that commute with $L_{s}$~\cite{loveridge2011measurement}.

The WAY theorem is not particularly intuitive.  If one reviews the standard proofs of it, one finds that while they
are mathematically rather straightforward, there is no compelling narrative accompanying them.  One of our goals here is to provide conceptual clarity on the origin of the WAY no-go result.

We consider the WAY theorem from the perspective of the resource theory of asymmetry~\cite{gour2008resource,marvian2011theory,bartlett2007reference,bartlett2006degradation,vaccaro2008tradeoff,gisin1999spin,chiribella2004efficient} (as was done recently in Ref.~\cite{Ahmadi2012WAY}).
Using a duality relation that is evident within the resource theory perspective,
we show that the WAY theorem can be derived from a fundamental result in quantum information theory: the no-programming theorem for projective measurements~\cite{duvsek2002quantum,DAriano2005efficient}.
To \emph{program} a measurement is to perform one of a set of possible projective
measurements with the choice being specified by the quantum state of an
ancilla (the quantum program). The no-programming theorem states that one cannot program distinct projective measurements using non-orthogonal program states.

In addition to the `no-go' aspect of the WAY theorem, both Wigner as well as Araki and Yanase showed that
it is possible to \emph{approximate} a repeatable measurement of an observable
$O_{s}$ that does not commute with $L_{s}$ given an appropriate state of the bounded-size apparatus.
Subsequent work has shown that this also holds true for measurements that are not repeatable~\cite{loveridge2011measurement,Ahmadi2012WAY}.
This also follows in an intuitive manner from our result: it is known that despite the no-programming theorem one can achieve \emph{approximate} programming of projective measurements~\cite{duvsek2002quantum,DAriano2005efficient,Vidal2002Storing}.
Each such result implies a corresponding result in the WAY context.

By identifying the WAY theorem as a special case of the no-programming theorem for projective measurements, our letter contributes to the project of uncovering the logical relations that exist among important quantum
phenomena. We also show that we can replace the assumption of a conservation law in the WAY theorem with the assumption that the dynamics is symmetric under the action of some compact Lie group.  This reformulation leads to a generalization of the WAY no-go result to the case of finite and non-compact Lie groups.  We also discuss its generalization to the case of nonprojective measurements and to the case of unitaries.

WAY-type restrictions have practical significance in quantum information processing, where there may be size limits on the probe used to access quantum systems~\cite{bartlett2007reference,bartlett2006degradation}.  They are also significant in the field of quantum gravity, where gravitational effects place limits on the quantum numbers appearing in the states of physical system.  For instance, there is an upper bound on the energy content of a finite region of space-time determined by the physics of black holes, which implies, via the WAY theorem, that it is impossible to measure distances in space-time to arbitrary accuracy using a device of finite spatial extent~\cite{AmelinoCamelia1994}.


\textbf{Conservation laws versus symmetry.} Traditionally, the WAY\ theorem is based on the assumption of a conservation law, but the latter can always be understood to be a consequence of symmetries of the dynamics.
This suggests that one should take the assumption of symmetric dynamics to be the origin of the restriction implied by the WAY theorem. Although the symmetries of the dynamics are typically assumed to be axiomatic, one can just as well imagine that they are due to practical considerations.
For instance, if two parties fail to share a reference frame for some symmetry group, then any dynamical evolution relative to one party's frame is described relative to the other party's frame as a symmetric operation~\cite{bartlett2007reference}.  Such practical considerations also impose a constraint of symmetry on the measurements that can be implemented and the states that can be prepared relative to the other party's frame.
This restriction to symmetric operations is sometimes called a superselection rule.

Now consider an experimenter, Bob, who is restricted to symmetric operations. Suppose a friend who does not face the restriction, Alice, gives Bob access to a sample of a quantum state that breaks the
symmetry (in other words, an \emph{asymmetric }state) or gives him access to a
single use of an asymmetric measurement or a single use of an asymmetric
transformation.  In this case, Bob can simulate other single-shot
operations that break the symmetry (and therefore lie outside the restricted
set). In this sense, the single-shot operations gifted to Bob by Alice
can be considered to be \emph{resources}: he could not have implemented them
himself, and when he uses them to simulate other single-shot operations, they
are consumed or degraded in the process.  The mathematical theory
that describes the interconversion of asymmetric resources under symmetric
processing is known as the \emph{resource theory of asymmetry}~\cite{bartlett2007reference,gour2008resource,marvian2011theory,marvian2011pure}.

It is useful at this stage to be more precise about how to define this resource
theory for a given symmetry group. \ One specifies the symmetry
of interest by specifying an abstract group $G$ of transformations and the
appropriate projective representation thereof.\ For instance, if one is
considering rotational symmetry, then the group is SO(3), and the projective unitary
representation can be written as $\left\{  R_{\mathbf{\hat{n}}}\left(
\theta\right)  =e^{i\theta\mathbf{J\cdot\hat{n}}}:\theta\in\lbrack0,2\pi
),\hat{n}\in S_{2}\right\}  ,$ where $\mathbf{J}=\left(  J_{x},J_{y}%
,J_{z}\right)  $ is the vector of angular momentum operators.  Here,
$R_{\mathbf{\hat{n}}}\left(  \theta\right)$ rotates by an angle of $\theta$
around the $\mathbf{\hat{n}}$ axis.

Suppose $\mathcal{H}$ is a complex Hilbert space and $g \mapsto U\left(  g\right)  $ is
the projective unitary representation of $G$ on $\mathcal{H}$ corresponding to
the symmetry of interest (we will sometimes add a subscript to specify the system on which the symmetry acts). Suppose $\mathcal{L}\left(  \mathcal{H}\right)  $
is the set of linear operators on $\mathcal{H}.$ A state $\rho\in
\mathcal{L}\left(  \mathcal{H}\right)  $ (i.e. a trace-one positive operator)
is said to be symmetric if $\forall g\in G:U(g)\rho U^{\dag}(g)=\rho,$ and an
observable $O$ is said to be symmetric if $\forall g\in G:U(g)OU^{\dag}(g)=O.$
\ \ Similarly, a unitary operator $V$ representing a quantum dynamics is said
to be symmetric if $\forall g\in G:U(g)VU^{\dag}(g)=V.$ \ It follows that if $G$ is a Lie group, then
symmetric states, observables and unitaries commute with all the generators of $G$. \ For instance, the states,
observables and unitaries that are rotationally-symmetric are those that commute with $\mathbf{J}\cdot\mathbf{\hat{n}}$ for any direction $\mathbf{\hat{n}}$.

If a unitary evolution has the symmetry associated with the generator $L$, then clearly $L$ is conserved under the evolution.  Furthermore, the observables that commute with the conserved quantity $L$ are those that have the symmetry. Those that \emph{fail} to commute with $L$ (which are the ones of interest in the WAY theorem) are those
that \emph{break} the associated symmetry.

In the original WAY analysis, the initial state of the apparatus is allowed to not commute with the conserved quantity, which means that it is allowed to break the symmetry. The alternative to this assumption ---a \emph{symmetric} initial apparatus state--- would trivially yield a no-go result because we require \emph{some} resource of asymmetry if we are to simulate an asymmetric measurement.
Therefore, from the perspective of the resource theory of asymmetry, the problem that the WAY theorem addresses is whether it is possible to perfectly simulate an asymmetric measurement given the resource of an asymmetric state using only operations that are symmetric.

Any such simulation can be achieved with a circuit of the following form: the system $s$ (upon which the measurement is to be implemented) and an ancilla $a$ (which contains the resource of asymmetry and possibly some additional system prepared in a symmetric state) are subjected to a joint measurement of the symmetric observable $\tilde{O}^{sa}$. Note that any symmetric unitary on $sa$ implemented prior to the measurement can always be absorbed into the definition of the measurement. Also, no generality is lost by not considering symmetric positive operator-valued measures (POVMs) on the composite because these can always be simulated by injecting an additional ancillary system prepared in a symmetric state and implementing a symmetric projective measurement on the whole (this follows from a covariant version of the Stinespring dilation theorem~\cite{Keyl1999Optimal}).   Note that, unlike the original WAY treatment, we have not assumed that the measurement is repeatable.

If the asymmetric resource state is $\rho_a \in\mathcal{L}\left(  \mathcal{H}
_{\mathrm{a}}\right)$ and one seeks to simulate a measurement of the asymmetric observable $O_s$ (with spectral projectors $\{ \Pi^{(s)}_{k}$\}),  then the question addressed by the WAY theorem is
whether it is possible to find a symmetric observable on the composite, $\tilde{O}_{sa}$ (with symmetric spectral projectors $\{ \tilde{\Pi}_{k}^{(sa)} \}$),
such that for all states of the system $\rho_s \in\mathcal{L}\left(
\mathcal{H}_{\mathrm{s}}\right),$ and for all outcomes $k$,
\[
\mathrm{Tr}_{sa}\left(  \tilde{\Pi}_{k}^{(sa)} \left(  \rho_s \otimes\rho_a\right) \right)
=\mathrm{Tr}_{s}\left(  \Pi^{(s)}_{k} \rho_s \right)  .
\]
If such a simulation exists, we say that $\rho_a$ is mapped to $O_s$ by symmetric processing and we write $\rho_a \xrightarrow{\textrm{sym}} O_s$.

To prove our main result, we must introduce a second kind of simulation problem: each element of the set of asymmetric states $\{ U_a(g) \rho_a U_a^{\dag}(g):g\in G\}$ must simulate the corresponding element of the set of asymmetric observables $\{ U_s(g) O_s U_s^{\dag}(g):g\in G\}$ but the processing that achieves the simulation can be arbitrary (i.e. it need not be symmetric).
Specifically, the question of interest in this simulation problem is whether it is possible to find some observable $O_{sa}$ on the composite (not necessarily symmetric)
such that for all states of the system $\rho_s \in\mathcal{L}\left(
\mathcal{H}_{\mathrm{s}}\right),$ and for all outcomes $k$, we have
\begin{align}
\forall g\in G &:\mathrm{Tr}_{sa}\left(  \Pi_{k}^{(sa)} \left[  \rho_s \otimes U_a(g) \rho_a U_a^{\dag}(g) \right] \right)  \nonumber \\
&=\mathrm{Tr}_{s}\left( U_s(g) \Pi^{(s)}_{k}U_s^{\dag}(g)\; \rho_s \right),
\end{align}
where $\Pi_{k}^{(sa)}$ are the spectral projectors of $O_{sa}$. If such a simulation exists,
we write $\forall g \in G: U_a(g) \rho_a U_a^{\dag}(g) \rightarrow U_s(g) O_s U_s^{\dag}(g)$.

We can now state the critical lemma from which our result follows.
\begin{lemma}\label{lemma:duality}
The following conditions are equivalent:
\item (i) $\rho_a \xrightarrow{sym} O_s$
\item (ii) $\forall g \in G: U_a(g) \rho_a U_a^{\dag}(g) \rightarrow U_s(g) O_s U_s^{\dag}(g)$
\end{lemma}

This lemma, which we shall prove shortly, demonstrates the existence of a
duality of perspectives on our resource interconversion problem. \ In version
(i) of the problem, the question is whether one resource, a single copy of the
state $\rho_a,$ can be transformed to another, a single implementation of the measurement of observable $O_s$, via operations that are restricted to be symmetric. This is the perspective of \emph{restricted dynamics}.  On the other hand, in version (ii) of the problem, the question is whether a single copy of the state $U_a(g)\rho_a U_a^{\dag}(g)$ can be transformed to a single implementation of the measurement of observable $ U_s(g)O_s U_s^{\dag}(g)$ via any operation (i.e., there is no longer any restriction to symmetric operations) but where the operation must achieve the conversion for \emph{every} value of $g.$ Given that a sample of the state $U_a(g)\rho_a U_a^{\dag}(g)$ encodes information about the group element $g,$ and a single implementation of the measurement of observable $\left\{  U_s(g)O_s U_s^{\dag}(g)\right\}  $ also encodes information about $g$ (one can derive an estimate of $g$ by implementing the measurement on a known state), version (ii) of the interconversion problem asks whether we can transform one encoding of $g$ into another and hence provides an \emph{information-theoretic} perspective on the problem.

Because one of our goals is to make the WAY\ theorem particularly intuitive, we present a proof of the lemma that appeals to a simple physical argument.

Imagine two parties, Alice and
Bob, each of whom has a local reference frame, but where these are related by an unknown group
element.
Alice prepares a system $a$ in the state $\rho_a$ relative to her local reference frame and sends it, along
with a classical description of $\rho_a$, to Bob. She also sends him a classical
description of a Hermitian operator $O_s$, and asks him to use the sample of
$\rho_a$ to implement a measurement on $s$ of the observable that is described by $O_s$ relative to her frame.
As an example, Alice may ask Bob to implement a measurement of spin along her $\hat{z}$-axis using a spin 1/2 system prepared in an eigenstate of $\hat{z}$-spin as a token of her $\hat{z}$-axis.
Given his lack of knowledge of the relative orientation of their frames, the only operations that Bob can implement relative to Alice's frame are the symmetric ones (for a proof of this, see Ref.~\cite{bartlett2007reference}).
Statement (i) of lemma~\ref{lemma:duality} is simply the description of the interconversion problem relative to Alice's frame.

The lemma is proven by noting that statement (ii) describes the same interconversion problem but this time relative to Bob's frame. Suppose $g$ is the unknown group element that relates Alice's frame to Bob's.
The system $a$ that Bob receives is described relative to his frame by the quantum state $U_a\left(  g\right)  \rho_a
U_a^{\dag}\left(  g\right)  ,$ and his task is to use it to simulate a
measurement of the observable on $s$ that is described relative to his frame by $U_s\left(  g\right)  O_s U_s^{\dag}\left(  g\right).$ \ Given that he can implement \emph{any } processing relative to his own frame, there is no restriction on the operations he can perform.
Because $g$ is unknown, however, the processing must work for every $g\in G.$

From the duality exhibited in the lemma, it is clear that one can characterize the asymmetry properties of a quantum state $\rho$ by the information-theoretic properties of the encoding of the group that it provides, namely, the encoding $\{ U(g)\rho U^{\dag}(g) \}$~\cite{marvian2011theory}.  We will call an encoding of a message \emph{perfectly informative} if the quantum states representing distinct possibilities for the message are all mutually orthogonal.  Because a classical encoding is one wherein the quantum states are mutually commuting, a perfectly informative encoding is always classical. In a similar way, we can define a state $\rho$ to be \emph{perfectly asymmetric} if the elements of the group orbit of $\rho$ are all mutually orthogonal.  If we define a state to have classical asymmetry if the elements of its group orbit are mutually commuting, then a perfectly asymmetric state is necessarily one with classical asymmetry. We will see that what determines whether or not a WAY-type no-go result applies is whether or not the resource state is perfectly asymmetric.

\textbf{The no-programming theorem.}
The no-programming theorem states that one cannot program distinct projective measurements using non-orthogonal program states; the program must contain a perfectly informative encoding of the identity of the projective measurement. The impossibility claim was first made in Ref.~\cite{duvsek2002quantum} and
a proof was provided in Ref.~\cite{DAriano2005efficient}.
We will present a new proof of this result that is particularly intuitive and easily generalized to a derivation of bounds on the accuracy with which one can achieve approximate programming.

Suppose we seek to program a device to implement a measurement of either $\hat{z}$-spin or $\hat{x}$-spin on a spin-1/2 particle.  Assume that the associated program states are $|\phi(\hat{z})\rangle$ and $|\phi(\hat{x})\rangle$ respectively. (There is no loss of generality in taking the program states to be pure because a
mixed state can always be purified using a larger ancilla space.)  Programmability implies that there is a device that maps both $|{+\hat{z}}\rangle |\phi(\hat{z})\rangle$ and $|{+\hat{x}}\rangle |\phi(\hat{x})\rangle$ to the ``{+}'' outcome, and maps both $|{-\hat{z}}\rangle |\phi(\hat{z})\rangle$ and $|{-\hat{x}}\rangle |\phi(\hat{x})\rangle$ to the ``{-}'' outcome.  Therefore, such a device can perfectly discriminate $|{+\hat{z}}\rangle |\phi(\hat{z})\rangle$ and $|{-\hat{x}}\rangle |\phi(\hat{x})\rangle$.  Given that $|{+\hat{z}}\rangle$ and $|{-\hat{x}}\rangle$ are non-orthogonal, it follows that $|\phi(\hat{z})\rangle$ and $|\phi(\hat{x})\rangle$ must be orthogonal.

It is clear that the same sort of argument applies for \emph{any} two distinct observables, associated with measurements having any number of outcomes. It follows that the dimension of the program space must be at least as large as the number of distinct projective measurements to be programmed.
Furthermore, the degree of distinguishability of nonorthogonal program states puts an upper bound on the accuracy with which one can achieve an approximate measurement of the target observable.

It is now straightforward to prove our version of the WAY theorem.  We have shown that the WAY theorem concerns the task of perfectly simulating a measurement of an asymmetric observable $O_s$ using an asymmetric state $\rho_a$ under symmetric processing.  By lemma~\ref{lemma:duality}, this is seen to be equivalent to the task of performing one measurement from the group orbit of $O_s$, using a quantum program prepared in the corresponding element of the group orbit of $\rho_a$.  But the no-programming theorem asserts that we can only succeed in this task if the quantum program provides a perfectly informative encoding of the target measurement, and this implies that we can only simulate the measurement of $O_s$ if $\rho_a$ is perfectly asymmetric.

In particular, if the group has an infinite number of elements, as is the case for any Lie group, while $\rho_a$ lives in a finite-dimensional space, then $\rho_a$ cannot be perfectly asymmetric and hence perfect simulation is impossible.  We can also easily recover the fact that no resource of asymmetry is required to simulate $O_s$ if the latter is a symmetric observable because in this case the group orbit of $O_s$ contains only a single element.

\textbf{Discussion.}
We have clarified the assumptions from which a WAY no-go result may be derived: whether or not the resource state is perfectly asymmetric (i.e. whether the elements of its group orbit are mutually orthogonal) determines whether or not it can be used to achieve a perfect simulation of an asymmetric observable.  Constraints on the `size' of the resource system, which have been the focus of previous discussions, can be easily derived from this condition.
Furthermore, armed with our clarified statement of the assumptions, we can obtain several generalizations of the WAY theorem.

For one, we can generalize the result to the case of finite groups.  Unlike Lie groups, there are no generators of the group action in this case, and so the traditional WAY analysis, which focusses on conservation laws, is not applicable. Nonetheless, our result implies that a perfect simulation of an asymmetric projective measurement is impossible whenever the dimension of the Hilbert space of the resource state is less than the order of the group.

We can also treat the case of noncompact groups.  For instance, suppose the group of interest is the Heisenberg-Weyl group, corresponding to the translations and momentum boosts of a nonrelativistic particle.  If under processings that are symmetric with respect to this group, one seeks to implement an asymmetric projective measurement, then the state of the asymmetric resource state must be mapped to an orthogonal state by every combination of a spatial translation and a momentum boost.  This is clearly impossible if the resource system consists of a single particle, but if it consists of a \emph{pair} of particles, then (leaving aside nuances concerning how to rigorously treat improper eigenstates of position and momentum) it is possible.  For instance, one of the particles could be prepared in an eigenstate of position and the other in an eigenstate of momentum.

Our analysis can also be applied to determining the extent to which one can approximate the measurement of an asymmetric observable using a state that is not perfectly symmetric.
Given that processing cannot increase the amount of information in an encoding, the group orbit of the resource state must encode at least as much information about the group, according to any measure, as the group orbit of the target observable. Such measures can be interpreted as measures of the asymmetry of the state or observable. Measures of asymmetry are therefore expected to be a useful tool for deriving good quantitative versions of the WAY theorem.

The original WAY result does not apply to generalized measurements, i.e. those associated with a nonprojective POVM. By adopting the perspective of the resource theory of asymmetry, one can easily infer that certain asymmetric nonprojective POVMs also suffer a WAY-type no-go result, as demonstrated in Ref.~\cite{Ahmadi2012WAY}.  Other POVMs are not subject to the no-go result, however, such as the approximate versions of projective measurements that can be simulated by resource states that are not perfectly symmetric.  This prompts the following question: what is the set of asymmetric POVMs that cannot be implemented when the available resource state is not perfectly asymmetric?  By our duality result, this problem is equivalent to the following one: for which POVMs is it the case that the group orbit of the POVM cannot be programmed unless the program provides a perfectly informative encoding of the group?  We conjecture that it includes the set of POVMs that are extremal relative to quantum pre-processing~\cite{Buscemi2005Clean}.

It is well-known that there is also a no-programming theorem for unitaries in quantum theory: if we are to implement one of a set of unitaries by programming, then the program must supply a perfectly informative encoding of the target unitary~\cite{Nielsen1997Programmable}. By an argument analogous to the one presented in this letter, this result implies the impossibility of simulating an asymmetric unitary unless one has access to a resource state that is perfectly asymmetric.  Therefore, there is a WAY-type no-go result for asymmetric unitaries.

Note that the conditions for simulating asymmetric observables are the same as the conditions for simulating asymmetric unitaries, namely, that the resource state has perfect asymmetry.  Such a resource state can be cloned by symmetric processing (because the elements of its group orbit are mutually orthogonal) and it can be shown that any other asymmetric state can be obtained from a perfectly asymmetric state by symmetric processing.
It follows that a state that is perfectly asymmetric allows for the simulation of any asymmetric state, measurement or transformation, any number of times. Therefore, it completely lifts the restriction to symmetric operations (the superselection rule) which defines the resource theory.  In other words, it constitutes a \emph{perfect reference frame} for the symmetry group of interest.  Therefore, the circumstances under which WAY-type restrictions do not apply are precisely those wherein asymmetry is no longer considered a resource.

\section{Acknowledgements}
RWS thanks Howard Wiseman and David Jennings for comments on a draft of this article.
Research at Perimeter Institute is supported in part by the Government of Canada through NSERC and by the Province of
Ontario through MRI. IM acknowledges support from NSERC, a Mike and Ophelia Lazaridis fellowship, and an ARO MURI grant.

\bibliographystyle{apsrev4-1}
\bibliography{WAYTheorem}

\end{document}